# The physical condition from the perspective of complexity: Application of a physical activity program for Alzheimer's disease.

Y. de Saá Guerra[1], J. M. Martín González[2], S. Sarmiento Montesdeoca[1], D. Rodríguez Ruiz[1], D. Rodríguez Matoso[1].

[1]Department of Physical Education. University of Las Palmas de Gran Canaria.
[2]Department of Physics. University of Las Palmas de Gran Canaria.

Corresponding author: yvesdesaa@gmail.com (Y. de Saá)

## Abstract

Alzheimer's disease is a sickness that has been studied from various areas of knowledge (biomarkers, brain structure, behavior, cognitive impairment). Our aim was to develop and to apply a protocol of programmed physical activity according to a non-linear methodology to enhance or diminish the deterioration of cognitive and motor functions of adults with this disease, using concepts of complexity theory for planning and programing the program. We evaluated 18 patients (12 women and 6 men) diagnosed with mild and moderate grade. We worked in small groups (2 people/coach), a total of 16 weeks, 5 sessions per week and 60 minutes per session. We designed a protocol that integrated work balance, joint mobility, coordination, muscular strength and metabolic efficiency. The strong point of this intervention was how to design the planning of tasks for patients. The tasks were developed in parallel to processes of evocation and cognitive association individually designed and presented in order of complexity. In all subjects studied, we found an improvement in the results obtained in the Mini-Mental State Examination screening test, test of Cricthon, Set-Test of Isaacs, Clock Drawing Test and Trail Making Test, and the evaluation test of fitness. The physical activity program designed and the methodology used, has been an adecuated tool to improve quality of life and autonomy of subjects analyzed.

## Introduction

The human brain and nervous system as a whole is, from the morphological and functional viewpoint, a similar structure to a complex network (Vicsek, 2002; Amaral and Ottino, 2004; Solé, 2009; Scheffer et al., 2009). It is a fact that its architecture generates behaviors that are impossible to understand from the individual study of its components (neurons). Some techniques such as electroencephalography (EEG) (Micheloyannis et al., 2006; Stam et al., 2007) and functional magnetic resonance imaging (FMRI) (Egúíluz et al., 2005; Achard et al., 2006) have been used to treat to decipher the brain structures and the underlying design. The brain network constructed based on these functional measures are called the functional network. More recently, structural measurements, such as cortical thickness, were also used to build the anatomical network (He et al., 2007).

The Alzheimer's disease (AD) is a neurodegenerative disease characterized by or accumulation of proteins such as Tau or β-amyloid, which causes progressive dementia in adulthood, leading to a state of total disability, and death in a period usually less than two decades (Robles et al., 2002). The congnitive reserve (CR) represents the cognitive and intellectual ability that a person has accumulated throughout his life. The cerebral reserve hypothesis assumes that both the innate intelligence and life experiences (education, work, health, habits) may only provide a reserve in the form of cognitive skills that enable better tolerate neuronal pathological changes.

Studies in humans suggest that physical activity may reduce the risk of cognitive impairment (Weuve et al., 2004). Authors as Abbot (2004), Dishman (2006) and Weuve (2004), have conducted studies with older people seeking the cognitive benefits that physical activity produces. They concluded that physical activity has a positive effects, not only physiological, but it also help to improve cognitive and psychological needs of older people. The participation in cognitively stimulating activities contribute to cognitive reserve (CR) (Wilson et al., 2003). If the cognitive reserve is based on the level of efficiency and flexibility of cognitive systems, it seems likely that the frequent use of these systems in tasks involving an intellectual challenge would be associated with higher level of cognitive reserve (Stern, 2002). In a study by Zabar et al. (1996), which evaluated the time that a group of people dedicated to participate in complex activities (playing a musical instrument, doing crafts) and basic activities (eating, dressing, and the like), the results showed that regular participation in complex activities versus the simple tasks reduces in two years the risk of developing dementia.

In the same line, Fabrigoule et al. (1995) noted that activities such as travel, complex work and weave were associated with a lower risk of dementia. Those seniors who have more leisure activities have a 38% lower risk of developing dementia, and the risk is reduced about 12% for each leisure activity adopted (Scarmeas et al., 2001). In another study also found that the frequency of reading newspapers, magazines or books was associated with a reduced risk of AD by 33% (Wilson et al., 2002). And as for the subjects with AD, who read more before develop the disease, showed less decline in verbal skills and general cognitive status compared with those who had little reading habits (Wilson et al., 2002).

Some evidence suggests that there may be further beneficial effects of physical activity on cognition throughout life, although the optimal amount and type of physical activity to support optimal brain function, probably vary over the life of people (Gracia and Marcó, 2000). It can reduce the risk of cognitive impairment and dementia by improving CR (Fabrigoule et al. 1995; Laurin et al. 2001; Scarmeas et al. 2001; Wang et al. 2002; Wilson et al. 2002). Several papers have highlighted the role of physical activity programs enriched with cognitive tasks in the stimulation of the psychological functions of elderly people (Peña-Casanova, 1999; Pont, 2007). These programs not only estimulate memory, but also emphasize on the rest of cognitive skills (Pont, 2007). They are methods of activities involving intentional movement encouraging intelligent-motor cognitive skills of attention and short-term memory, perceptual-motor skills of balance, spatial and temporal structure, coordination and body awareness, physical abilities conditional joint mobility, muscle flexibility and endurance and social skills related to interpersonal communication (Pont, 2007).

The aim of the study was to determine whether planning a physical activity protocol, developed from the methodological perspective of complex systems, can improve the physical and cognitive impairment cushion involved in Alzheimer's disease.

## Physical Activity improve cognition

One of the mechanisms by which physical activity may be beneficial for cognition is that it stimulates trophic factors and neuronal growth, possibly providing a reserve against degeneration and dementia (Van Praag *et al.*, 1999).

Has also been shown that physical activity maintains and stimulates the cerebral blood flow (Rogers et al., 1999) by increasing the vascularity of the brain (Dik et al., 2003). This can cause an improvement in aerobic capacity and the arrival of nutrients to the brain (Spirduso 1980, Dustman et al., 1984). In addition to structural changes in the brain (brain reserve), physical activity early in life may increase the functional capacity of the brain (cognitive reserve), increasing nerve efficiency (Stern, 2002). In many cases be prescribed aerobic activity as a means to improve breathing capacity and the greatest means of protection as a result of cognitive functions in as (Barnes et al., 2003).

Dishman (2006), suggests that physical activity can provide protection and benefits for several neurological diseases including Parkinson's disease, Alzheimer's dementia, and ischemic stroke. The author asserts that physical activity improves health in cognitive function in older people and has a beneficial effect on depression and sleep quality. Recently, he has accumulated convincing evidence that physical activity and regular exercise can reduce depression, cognitive impairment associated with aging, and provide protection from disease, trauma and brain injury by influences on axonal growth and synaptic plasticity.

Weuve (2004), maintains that physical activity programs in older people, provide an unquestionable improvement in both memory as in other collateral cognitive factors (spatial concept, visual reasoning, orientation, emotional, personal and social issues). Their prospective study concludes that regular physical activity, long duration (aerobic), is strongly associated with less cognitive impairment and better cognitive performance among older people.

| Influence of physical activity in Alzheimer's disease |
|---|
| **Overall reduction of risk factors:**<br><br>Reduction of cardiovascular risk factors: hypertension, glucose tolerance, insulin resistance, lipid profile, overweight<br>Reducing the risk of cerebrovascular accident<br>Improves cerebral blood flow and oxygen supply<br>Promotion of endothelial nitric oxide production<br>Reduction of inflammation<br>Decreased accumulation of oxidative radicals<br>Promotion of brain plasticity<br>The increase cognitive reserve<br>Increased social activity |
| **Increased cerebral cytoarchitecture:**<br><br>The increase in dendritic length, the proliferation of neuronal progenitors, dendritic complexity<br>Growth of blood vessels in the hippocampus<br>Growth of blood vessels in the cortex<br>Growth of blood vessels in the cerebellum<br>The increase in microglia<br>Increased short and long term potentiation in the dentate gyrus<br>The increase in cerebral capillary density<br>Promoting the expansion of nerve fibers<br>Proliferation of microglia in the cortex<br>Increased neurogenesis and proliferation<br>The reduction of brain tissue loss in the hippocampus<br>Increased number of differentiating neurons |
| **Improving the electrophysiological properties:**<br><br>Enhancing reinforced in response to high frequency stimulation<br>Increased level of synapsin and synaptotrophin<br>Increased glutamate receptors (NR2B and GluR5) |
| **The increase in brain growth factors:**<br><br>Increased brain-derived neurotrophic factor.<br>Down-regulation of hippocampal neurotrophin (NT3), BDNF and neurogenesis in the hippocampus |
| Increased insulin-like growth factor-1 (IGF-1) |
| Increased endothelium-derived growth factor (VEGF) |
| Increased serotonin |
| The increase in acetylcholine |
| Induce fibroblast growth factor |
| **Impact on the burden of amyloid:** |
| Reduction of amyloid burden |
| Stabilized amyloid burden |
| Improvement of the hippocampus, despite high APP |
| Increased amyloid burden |
| **Other mechanisms:** |
| Hippocampus increased expression of enzymes involved in glucose utilization |
| Change in gene transcription |
| Increased level of calcium in the central nervous system |

Table 1. Main conclusions in biological research on the enriched environment (physical activity) of Alzheimer's disease (extracted from Rolland et al., 2008).

## Methodology

*Sample*: we studied 18 patients (12 women y 6 men) diagnosed with mild and moderate AD (age: 75,78 ± 5,53 years [63-85]; body weight: 64,19 ± 13,15 Kg [42-89]; height: 1,57 ± 0,08 m [1,46-1,70]; BMI: 25,84 ± 4,28 Kg/m2 [18,92-34,44]). The sample belonged to a group of subjects of Canary Azheimer Association, who voluntarily participated in the intervention program. All kin of patients and the Director of Canary Azheimer Association were informed of the characteristics of the study and signed, as Tutors, a written consent. All study participants rigorously followed the criteria proposed in the Declaration of Helsinki for research involving human subjects.

*Evaluation proceeding*: Subjects were evaluated before, during and after the intervention program. We consider the subject as physically active subjects, because in this association, they received a daily program (small groups) of basic psychomotor stimulation (joint mobility, breathing exercises, drills and exercises of lateral coordinative) and cognitive stimulation for at least one year.

For the assessment of lower extremity strength, the main test designed was to measure the number of times they get up and sit in a chair in 30 seconds (*30 seconds chair stand*) (Rikli and Jones, 2001) and *foot up and go* test which measure the time they spend to cover a distance established (2,45m). We evaluated the effects of the program on the muscular response by Tensiomiography (TMG) (Dahmane et al., 2000; Belic et al., 2000, Simunic and Valencic, 2001, Simunic, 2003, Krizaj, 2008; Tous-Fajardo et al., 2010; Rodríguez-Matoso et al., 2010).

The cognitive assessment was done by different test of screening: MMSE (Mini-Mental State Examination) with maximum score of 30 points, Cricthon (indicates level of cognitive impairment) with maximum score of 38, STI (Set-Test of Isaacs) of verbal fluency with maximum score of 40, CDT or Clock Drawing Test with a maximum score of 10 points and TMT or Trail Making Test in parts (A/B) which collects the runtime. For the analysis of the data were taken as control variables to study the age, sex and years of schooling of the subjects in the sample.

*Intervention Protocol*: The intervention program for this study was conducted in small groups of 2 people per coach, during 16 weeks with a frequency of 5 sessions per week and sessions last for at least 60 minutes per session. We designed a circuit of stations where the subjects performed exercises of various guidelines aimed to develop balance, joint mobility, coordination, muscular strength and metabolic efficiency with the following protocol:

| **Orientation training (Objectives)** | | | |
|---|---|---|---|
| **Monday** | Muscular Strength | Coordination | Motricity - Associative- Evocative |
| **Tuesday** | Metabolic Efficiency | Mobility-Balance | |
| **Wednesday** | Muscular Strength | Coordinación | |
| **Thursday** | Metabolic Efficiency | Mobility-Balance | |
| **Friday** | Muscular Strength | Coordination | |
| **Saturday** | Motricity | Affective Socialization | |
| **Sunday** | Family Activities | Unifying | |

Table 2. Periodization and objectives.

**Coordination:** tasks involving fine motor manipulations: throwing and receiving of objects (one and two hands), manipulating objects, use of implements. And variants of these exercises.

**Balance:** tasks that involve improving the static and dynamic balance, balance shaft bipodal hip displacement (oscillations) bipodal balance by manipulation, changing the method of balancing. Unipodal balance, balance unipodal with manipulation. Movements in different directions and rhythms (open and close eyes); travel with address changes and rhythms, movements directed and conditioned. And similar exercises for the same purpose.

**Mobility**: tasks that involve improving of motion range of joints of upper and lower limbs. Stretching the muscles that affect the ankle, knee and hip. Stretching the muscles that affects the joints of the wrist, elbow and shoulder.

**Endurance** (metabolic efficiency), aerobic loads of low-impact and intensity at reclining cycleergometer adding fine motor and cooperatives tasks.

**Strength** (muscle tone): directed tasks, primarily, to the lower limb muscles with your own body weight or using external loads on weights machines.

All these tasks were developed in conjunction with processes of cognitive activation and associative evocation, individually designed and presented in order of complexity of the task. All coaches know the dossier and history of each subject in order to avoid those topics of conversation that could lead to increased levels of stress or a nervous breakdown.

The protocol would be incomplete without the application of additional stimulation to perform the other two days of the week. Its aim involves stimulation through activities (60') in different environments (natural environment) and family involvement in additional integration tasks related to joint physical activity and adequate level of affection (45´).

*Statistical Analysis.* For a description of the study variables of the sample was carried out descriptive statistics of the variables evaluated in the physical test (get up and sit, stand and walk and TMG) and cognitive [MMSE, CRICHTON, STI, CDT, TMT (A/B)].

After a normality análisis Shapiro-Wilk of the variables, we carried out a mean comparison between the results of the Pre-Intervention and the Post-Intervention for paired samples. A T-test calculation was used for comparison of means related parametric and T-Wilcoxon for comparison of means related nonparametric. The decision statistics were calculated by taking a level of significance $p<0,05$. For the statistical software we used SPSS-v17 (SPSS Inc., Chicago, IL, USA).

## Holistic training. Training based on complexity.

The most notably aspect of this intervention was the way to develop the working sessions, ie, the approach of how to conduct training sessions. The purpose of training was to achieve harmony among all the systems that compose the body and and how they relate to each other. It is when a set of elements can achieve behave similar and consistent (Watts and Strogatz, 1998).

We must keep in mind that each individual is unique, is the result of their genetics and their experiences. Therefore we must regard them when we plan implementing training activities. Siff (1999) emphasizes the need to identify, consider the style of each individual without to emphasize an ideal model, to diversify, to respect the asymmetry, the delayed effects and the interactive processes. That is, to adapt to what we are facing. Do not work in a completely analytical way, but be clear about the goal we want to get and use the necessary resources to achieve them. Thus the radical unpredictability is the creative foundation of the process. We always looking for the transition from a state of greater symmetry (less complex), to a lower symmetry (more complex).

Our approach was always respecting one of the fundamental principles of training: the principle of individuality of the loading, and with due regard to our overall objective, was to improve the system and not only one of its agents. A holistic individual processes multiple elements simultaneously to organize them into a complex unit, this is performance. A synergy.

The development of the session was carried out following a methodological approach increasing the complexity of the tasks that, in our case, have been adapted to the daily activities of everyday of patients (eg: getting out of bed, tying shoes, take a shower, picking up or placing objects on shelves, and the like) always searching for motion control adaptations, maximum efficiency and minimum risk to their safety. The tasks range from: loads with low complexity coordinated elements, towards transfer loads and coordinative elements more complex; from loads with the possibility of multiple responses towards loads of features incorporated into daily actions. We always try to perform tasks with variations in implementation and the magnitude of the load.

The use of a methodology based training complexity is based on the following phenomena:

Training Methodology non-linear: The idea is to generate exercises or task at which we introduces specific random elements that require that the subject maintain an attitude of continual adaptation to be able to maintain the correct execution: for example, dribbling of ball (soccer, basketball, hockey ball, or the like). The coach says numbers with the fingers of a hand randomly and with the voice says other differents. The person who perform the dribbling has to do it without errors (have to create new resources to move forward) and say the numbers of the coach hand, who is moving for all the space randomly.

The key is to introduce controlled stimuli (physical and cognitive), unknown in advance by the individual, so that should devise one or more answers to resolve the task (nonlinear output). Thus, as the individual improves their performance, optimize their own subsystems. This process is related to improvements in structures and internal resources of a complex system: in an environment relatively random in which must react, and because of adaptation (fitness), he improves his performance. From the standpoint of modern theories of chaos and complex systems, could be related to the evolution of complex adaptive systems, fitness or adaptation to the environment, or even the phenomenon of stochastic resonance. We try to they perform the coordination and evocation tasks simultaneously with the highest degree that the subject can perform. Always close to making mistakes. At the boundary of the error. Our goal was increasing their thresholds of capacity of evocation and cognitive-associative activation , and therefore improve brain circuits linked.

These tasks, which modify in a controlled way the environment also aim to activate behaviors that can be atrophied or inhibited. Introducing unpredictability in the work encourages creativity in resolution of motor and intellectual problems. What can be understood as a phenomenon of emergency. Increasing the capacity to respond to external perturbations by the interaction of its internal elements, enhance the capabilities of self-organization of these.

One example is the shooting exercise with balls of different size, weight, air pressure, different touches (smoother or rougher) and variations of this type. This forces the subject to adapt to different environmental conditions and adjust his systems to succeed in the task. Coordination tasks are a practical example of the phenomena of synchronization of neural circuits.

In short: We tried to improve the brain system through the stimulation of different areas (cognitive, motor and memory) interconnected (brain network) simultaneously. Our goal was to enhance the relationship between them (and the possible activation of inhibited behavior) and ultimately slow the progression of the disease.

Every complex system has elements, which in one way or another, exchange information with each other through some medium. This information flow is generated by the constituent elements, and in turn change the status of the last in a logical circle that we can not break (Solé, 2009). Interactions between units in phyical, biological, technological, and social systems usually give rise to intricate networks with non-trivial structure, which critically affects the dynamics and properties of the system (Guimerá, 2007).

## Results

The data obtained from patients with Alzheimer's screening test measured by a pre and post intervention, show the following results (Table 3):

| TEST SCREENING | PRE | POST |
|---|---|---|
| **MMSE** | 21,00 (± 1,23) | 19,43 (± 1,85) |
| **CRICHTON** | 22,52 (± 2,08) | 18,22 (± 2,39) |
| **STI** | 21,75 (± 3,03) | 19,94 (± 2,83) |
| **CDT** | 4,26 (± 0,84) | 4,73 (± 1,01) |
| **TMT A** | 240,33 (± 41,28) | 166,37 (± 24,96) |
| **TMT B** | 363,41 (± 54,32) | 287,64 (± 30,67) |

Table 3: Mean and standard deviation of the results of the screening test Pre and Post Intervention.

The data obtained from screening test MMSE, STI, CDI and in the A part of TMT, assessments resulting from Pre and Post intervention, show no significant differences. However, data from the screening test CRICHTON y TMT part B show statistically significant differences (p = 0,022 y p = 0,018 respectively). The data obtained in subjects on the parameters analyzed with the TMG in the post-intervention test show a positive change with respect to pre-intervention data for the speed response normalized (Srn) and maximum radial displacement of the muscle belly (Md) in each of the muscles analyzed (Table 4):

| Muscle | Srn (mm/s) | | | |
|---|---|---|---|---|
| | **Pre** | **Post** | **Differ.** | **Estad.** |
| **RF right** | 24.3 (± 3.8) | 34.0 (± 6.0) | 9.7 ** | *p*=0.000 *TE*=2.54 |
| **RF left** | 25.0 (± 4.6) | 33.8 (± 6.2) | 8.9 ** | *p*=0.000 *TE*=1.91 |
| **VL right** | 29.9 (± 7.7) | 31.04 (± 5.0) | 1.2 | *p*=0.435 *TE*=0.15 |
| **VL left** | 32.5 (± 8.2) | 30.2 (± 9.1) | - 2.3 | *p*=0.227 *TE*=-0.29 |
| **VM right** | 41.9 (± 12.7) | 41.2 (± 9.91) | - 0.7 | *p*=0.807 *TE*=-0.06 |
| **VM left** | 41.3 (± 9.6) | 37.2 (± 9.8) | - 7.7 | *p*=0.153 *TE*=-0.43 |
| **BF right** | 21.4 (± 8.0) | 39.9 (± 13.8) | 18.4 ** | *p*=0.000 *TE*=2.35 |
| **BF left** | 18.2 (± 9.3) | 40.8 (± 11.4) | 22.6 ** | *p*=0.001 *TE*=2.43 |

| Muscle | Md (mm) | | | |
|---|---|---|---|---|
| | **Pre** | **Post** | **Differ.** | **Estad.** |
| **RF right** | 3.2 (± 2.6) | 3.7 (± 2.3) | 0.5 | *p*=0.535 *TE*=0.20 |
| **RF left** | 4.1 (± 3.1) | 3.1 (± 1.5) | - 1 | *p*=0.180 *TE*= -0.31 |
| **VL right** | 3.6 (± 1.9) | 3.9 (± 1.3) | - 0.3 | *p*=0.435 *TE*=0.15 |
| **VL left** | 3.6 (± 1.8) | 3.2 (± 1.2) | 0.4 | *p*=0.381 *TE*= -0.24 |
| **VM right** | 4.2 (± 2.5) | 4.8 (± 2.4) | 0.6 | *p*=0.212 *TE*=0.24 |
| **VM left** | 4.2 (± 2.8) | 5.1 (± 2.7) | 0.9 | *p*=0.075 *TE*=0.33 |
| **BF right** | 4.2 (± 3.4) | 3.5 (± 2.3) | - 0.7 | *p*=0.811 *TE*= -0.23 |
| **BF left** | 3.5 (± 2.8) | 4.1 (± 3.2) | 0.6 | *p*=0.384 *TE*=0.22 |

Table 4: Mean and standard deviation of the results of Srn and Md for the muscles tested in Alzheimer's patients on the test before (Pre) and after (Post) the intervention, the resulting difference between test (** p ≤ 0.05) and the effect size (TE).

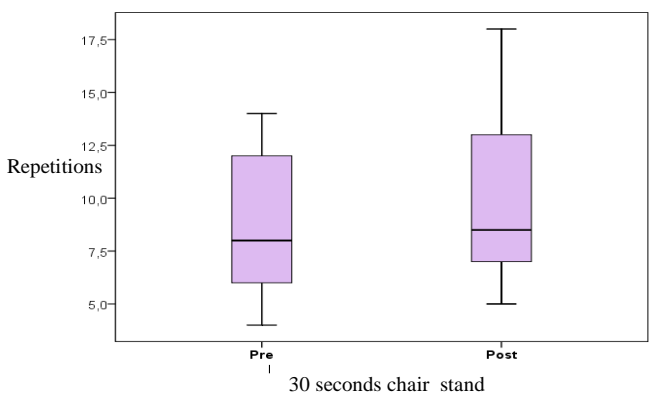

30 seconds chair stand

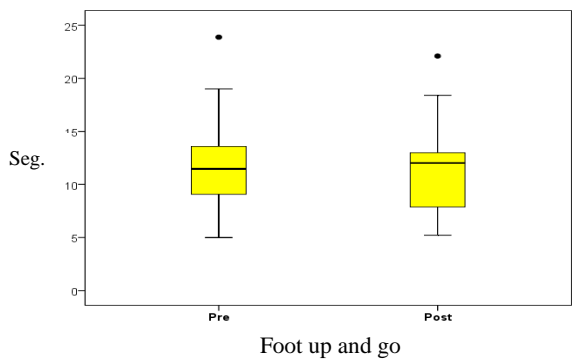

**Figure 1.** Box plot of the results of the test (above) and *foot up and go* (below).

In all subjects in our study we found a marked improvement in the results of evaluation tests of physical condition. In the *30 seconds chair stand* increased by 7.05% the number of repetitions performed after the program (PRE: 8,78 ± 3,46 vs. POS: 9,44 ± 3,68 repeticiones). In the *foot up and go* test (2,45m) decreased slightly (2,17%) the time which get up, cover the distance and sit down (PRE: 11,95 ± 5,19 vs. POS: 11,69 ± 4,43 seconds).

## Discussion

The results show that subjects with Alzheimer's dementia, which participated in a protocol of programmed physical activity (according to a non-linear methodology, based on complex systems behavior), manifest an improvement in their physical abilities, an increase of 7.05% in the repetitions in the *30 seconds chair stand* test, a decrease in time (2.17%) in the *foot up and go* test (2.45 m) and even an increase in normalized speed response rates (Srn) and the maximum radial displacement of the muscle belly (Md).

As well as a moderate reduction in cognitive impairment: greater than or equal number of hits on the MMSE, similar data in the Crichton, statistically significant differences (p=0.263) in the STI. In the Clock Drawing Test (CDT), without statistically significant differences. We affirm that there is no difference between pre and post valuations, which suggests a slowed disease progression, in terms of executive function. In Part-A of the TMT, we noted a remarkable decrease in execution time, justifying this trend (TE = 0,31) and in Part-B of the TMT, we can note a statistically significant difference (p <0.05) in regard to completion of the test time (time conditioned by the exeution).

It is possible that these improvements were due to the implementation of physical activity program. Whose objective was activating the motor control areas and motor action areas at the brain, while also stimulating brain areas associated with cognition, memory, evocative processes, perceptual and associative processes, areas associated with the development of discriminatory responses.

We believe this global vision is the way we should follow, because the system works in conjunction. Therefore it is very important choose well the working groups and the interaction of the coaches. We know that, unconsciously, our minds are connected. Create networks with people we have close through mechanisms that connect us with others, as is the case of mirror neurons (Iacoboni, 2009). It is possible that these mechanisms encourage the processes of our mental networks, and with an adequate training (stimulation) mental synchronization processes are positively enhanced. In this aspect perception plays an important role.

Perception is a bipolar process. According to James (1985) one part of what we perceive comes through the senses, what we have in front of us. And another part always comes from our own mind. Hence it be called bipolar process because of this dual side: the sensory and cognitive. Perception depends on the individual and personal characteristics of each subject. It depends on stimuli that act, the sensory organs. It depends, in short, the activity of perceiving subject and perceptual schemas. In the perception influence learning, past experiences, cultural, motivations, expectations, skills, personality, habits and memories. Perception is an active process. All influence on the selection and interpretation of data and content. Therefore, our training has taken a multidimensional approach.

## Conclusions

The conclusion of this work is that the subjects show a slight tendency of stabilization or a slowing of cognitive impairment aspects assessed in the MMSE, Crichton, STI and CDT in patients diagnosed with Alzheimer's dementia who participated in a specific physical activity program.

All subjects show a clear improvement of executive procedures by reducing the execution time of part A and B of the TMT. According to our results, we understand that participation in specific physical activity programs have a protective effect against the course of cognitive impairment, allowing an improved quality of life of patients diagnosed with Alzheimer.

All subjects improved physical abilities (strength, endurance, mobility and coordination), the muscle response (Tone and speed response), balance (static and dynamic) and mechanical properties and kinematics of the march. As well as a slight stabilization of cognitive aspects assessed in the MMSE, CRICHTON, STI and CDT.

The specific physical activity program proposed as methodology and designed to work with elderly diagnosed with Alzheimer's disease, was an effective tool for improving the quality of life and autonomy of subjects analyzed motor.

## Acknowledgments

To all subjects who participated in the study, their families, as well as Canary Alzheimer Association and all staff.